\documentclass[conference]{IEEEtran}
\usepackage[caption=false]{subfig}
\usepackage{cite}
\usepackage{amsmath,amssymb,amsfonts}
\usepackage{algorithmic}
\usepackage{graphicx}
\usepackage{textcomp}
\usepackage{multirow} 
\usepackage{array}
\usepackage{flushend}

\def\BibTeX{{\rm B\kern-.05em{\sc i\kern-.025em b}\kern-.08em
    T\kern-.1667em\lower.7ex\hbox{E}\kern-.125emX}}
\makeatletter
\newcommand\figref{Fig.~\ref}
\newcommand\secref{Section~\ref}
\newcommand\tabref{Table~\ref}
\begin{document}

\title{Estimation of 2D Velocity Model using Acoustic Signals and Convolutional Neural Networks }
\author{
\IEEEauthorblockN{
		Marco Paul E. Apolinario, 
		Samuel G. Huaman Bustamante,
		Giorgio Morales,
		Joel Telles,
		Daniel Diaz}
\IEEEauthorblockA{%
		National Institute of Research and Training in Telecommunications (INICTEL-UNI)\\
		National University of Engineering\\
		Email: mapolinariol@uni.pe}%}
}
\maketitle

\begin{abstract}

The parameters estimation of a system using indirect measurements over the same system is a problem that occurs in many fields of engineering, known as the inverse problem. It also happens in the field of underwater acoustic, especially in mediums that are not transparent enough. In those cases, shape identification of objects using only acoustic signals is a challenge because it is carried out with information of echoes that are produced by objects with different densities from that of the medium. In general, these echoes are difficult to understand since their information is usually noisy and redundant. In this paper, we propose a model of convolutional neural network with an Encoder-Decoder configuration to estimate both localization and shape of objects, which produce reflected signals. This model allows us to obtain a 2D velocity model. The model was trained with data generated by the finite-difference method, and it achieved a value of 98.58\% in the intersection over union metric 75.88\% in precision and 64.69\% in sensibility.\footnote{This paper is a preprint (submitted to the INTERCON 2019 conference, Lima, Peru). IEEE copyright notice. Ó 2019 IEEE. Personal use of this material is permitted. Permission from IEEE must be obtained for all other uses, in any current or future media, including reprinting/republishing this material for advertising or promotional purposes, creating new collective works, for resale or redistribution to servers or lists, or reuse of any copyrighted.}         

\end{abstract}


\section{Introduction}

The inverse problem consists of posing an approximate model of a system in which its parameters could be estimated from measurements, which are usually indirect. This kind of problem is present in multiple fields of science and engineering, moreover it is especially important in areas like geology and oceanography, where estimating soil properties or characteristics of underwater structures could only be possible through indirect measurements \cite{Linde, Collins}. Specifically, in the area of underwater acoustics, the estimation of position and shape of underwater objects is important for activities such as exploration and navigation, for that reason multiple approaches have been proposed, with analytical methods being the main ones \cite{Linde, Collins, Tarantola}. However, with the rise of high-performance computational methods such as deep neural networks, a great variety convolutional networks models have been proposed to solve inverse problems to different applications \cite{Moseley, Wu, Sorteberg}. 

For the above, in this work we propose a convolutional encoder-decoder architecture to estimate a 2D velocity model of an underwater environment, determining approximately the localization, shape and size of objects in the environment. The estimation of the velocity model is made from the reception of the echoes in 11 points of the study medium. To be able to train the convolutional neuronal network (CNN); a model, based on finite-differences method (FDM), is posed with which synthetic data are generated.

The remainder of this paper is as follows. In the \secref{sec:materials_methods}, we describe in detail the methodology used in this work, explaining the procedure to generate synthetic data and the architecture of the proposed CNN. Then, in the \secref{sec:results}, we present the results obtained in the training of the CNN and the analysis of CNN's performance. Finally, in the \secref{sec:conclusion}, we point out the conclusions obtained.

\section{Materials and Methods}\label{sec:materials_methods}

In this section, we describe the inverse problem studied here and, then, we present the methods used to generate data and estimate the solution of the inverse problem using convolutional neural networks.

\subsection{Inverse Problem}\label{sec:inversion_problem}

In the area of underwater acoustics, there is a variety of inverse problems, which, according to \cite{Collins}, are classified into two groups: remote sensing and source localization problems. In this paper we study a remote sensing problem, which seeks to estimate the location and shape of objects found in an aquatic environment; therefore, a CNN model is proposed that can perform this estimation using signals received from eleven points and produced by the propagation of an acoustic signal through the medium and from a point source.

\subsection{Synthetic Data Generation}\label{sec:dataset_generation}

We decided to use synthetic data to train the proposed CNN due to the fact that no public dataset with the required characteristics could be found, and also that obtaining real data would be very expensive in resources. For these reasons, we use the finite difference method to simulate the acoustic wave propagation in an underwater medium in order to generate 20,000 samples under different scenarios.

\subsubsection{Finite-difference method}\label{sec:finite_difference_method}

In order to generate synthetic data, we model the forward problem as the propagation of a plane acoustic wave in an 2D heterogeneous medium \cite{Yang-Hann, Lurton}. To this purpose, we solve the partial differential equation \eqref{eq:wave} using the finite-difference method under an extrapolation approach \cite{Heiner}; where $p$ is the pressure, $c$ is the wave propagation velocity in the medium, $s$ is the acoustic source, $x$ and $z$ are the spatial coordinates, and $t$ is the temporal coordinate.

\setlength{\arraycolsep}{0.0em}
\begin{eqnarray}\label{eq:wave}
\partial_{tt} p(x,z,t) & {}={} & c(x,z)^2 (\partial_{xx} p(x,z,t)+ \partial_{zz} p(x,z,t)) \nonumber\\
&&{+}\:s(x, z, t)
\end{eqnarray}
\setlength{\arraycolsep}{5pt}

The FDM requires to discretize the variables that are shown in \eqref{eq:wave} and approximate partial derivatives as a combination of these variables. To do this, the spatial coordinates $x$ and $z$ are divided into length spacings $dx$ and $dz$, while the temporal coordinate is segmented in time intervals $dt$. In \eqref{eq:index_label}, it can be seen the relation between continuous and discrete variables, where the superscript $n$ is related with timesteps, and subscripts $i$ and $k$ are related with $x$ and $z$, respectively .

\setlength{\arraycolsep}{0.0em}
\begin{equation}
\label{eq:index_label}
p_{i, k}^n = p(i dx,k dz,n dt)
\end{equation}
\begin{equation}
\label{eq:time_derivative}
\partial_{tt} p \approx (p_{i, k}^{n+1} - 2 p_{i, k}^n + p_{i, k}^{n-1})/dt^2
\end{equation}
\begin{eqnarray}\label{eq:x_derivative}
\partial_{xx} p & {}\approx{} &  ((p_{i+3, k}^n + p_{i-3, k}^n)/90-3(p_{i+2, k}^n + p_{i-2, k}^{n})/20\nonumber\\
&&{+}\:3(p_{i+1, k}^n + p_{i-1, k}^n)/20-49p_{i, k}^n/18)/dx^2
\end{eqnarray}
\begin{eqnarray}\label{eq:z_derivative}
\partial_{zz} p & {}\approx{} &  ((p_{i, k+3}^n + p_{i, k-3}^n)/90-3(p_{i, k+2}^n + p_{i, k-2}^n)/20\nonumber\\
&&{+}\:3(p_{i, k+1}^n + p_{i, k-1}^n)/20-49p_{i, k}^n/18)/dz^2
\end{eqnarray}
\setlength{\arraycolsep}{5pt}

In the approximation of the partial derivatives, we use three-point stencil for the temporal derivative and seven-point stencil for the spatial derivatives; their expressions are \eqref{eq:time_derivative}, \eqref{eq:x_derivative} and \eqref{eq:z_derivative}. From these equations, the variable $p$ is isolated at the timestep $n+1$ as a function of the previous timesteps, which are used to obtain the temporal evolution of the pressure.

\begin{figure}[tp]
\centering
\subfloat[]{%
        
        \includegraphics[width=0.4\columnwidth]{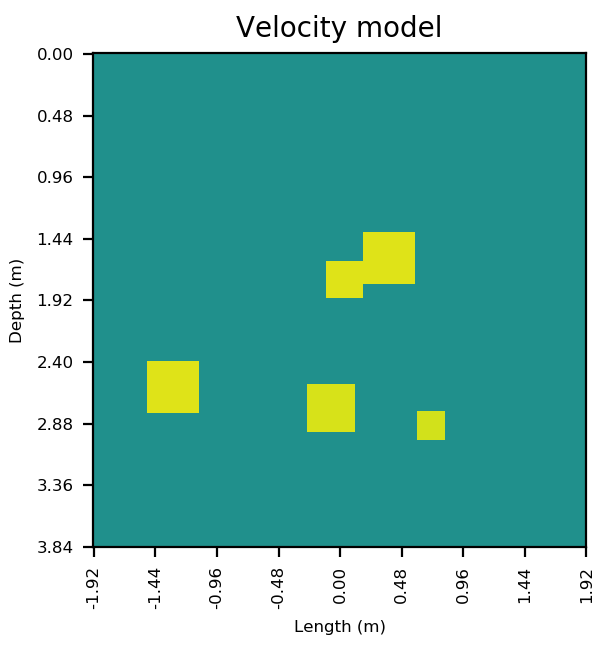}
        
        \label{subfig:vel_model}
}
\subfloat[]{%
        
        \includegraphics[width=0.4\columnwidth]{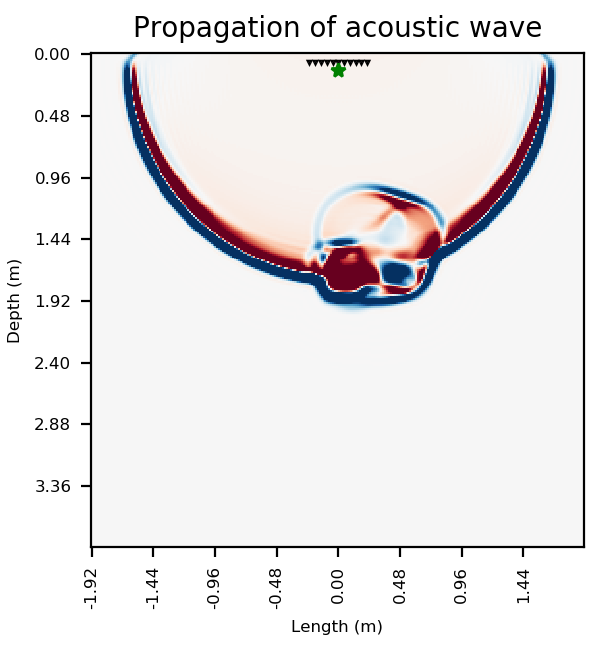}
        
        \label{subfig:wave_propagated}
}
\caption{ (a) Velocity model used in the simulation. (b) Propagation of the acoustic wave in the medium. Black inverted triangles represent measuring points while the green star represent the acoustic source.}
\label{fig:simulation}
\end{figure}

Then, we model the medium of propagation as a grid of points, where each point has a specific velocity, which is called the velocity model ($ c_ {i, k} $). By doing so, we can emulate the heterogeneity of the medium and build structures of different shapes within it so that we can also measure echoes produced by the structures when the wave impacts with them, as observed in \figref{subfig:wave_propagated}. The velocity values in the model are in the range of 0 m/s to 3000 m/s; in addition to this, we established the homogeneous medium as water with a velocity equal to 1500 m/s.

Moreover, the dimensions of the velocity model and variables $dx$ and $dz$ determine the physical dimensions of the simulated environment. Given that  $dx$ and $dz$ are equal to one-fifth of the minimum wavelength, that is 15 mm, and the model has a dimension equal to $256\times256$, the simulated space has a surface of 14.74 m2. Additionally, we considered a total of 1800 timesteps, each one with a duration of 2.5 $\mu s$, which was determined with the Courant-–Friedrichs-–Lewy condition \cite{Heiner}.  

Finally, an acoustic source is fixed in a point of the simulated space, as it is shown in \figref{subfig:wave_propagated}. This source produces a waveform equal to the first derivative of the Gaussian function \eqref{eq:signal_source}. We choose this function because it has a limited bandwidth. In \eqref{eq:signal_source}, the parameter $f_0$ is the maximum signal frequency, and it has a value of 40 kHz.

\begin{equation}
\label{eq:signal_source}
s^n = -2 (n-100) dt^2 f_0^2 e^{((n-100) dt f_0)^2}
\end{equation}

\subsubsection{Structure of samples}\label{sec:structure_samples}

Each individual sample in the dataset is an input-target pair, which is generated by the method described in \secref{sec:finite_difference_method} with different velocity models. We randomly generate velocity models (\figref{subfig:vel_model}) for each simulation, where each one has a different number of objects, in the range of 0 to 10. These objects are disk-or-square-shaped and randomly distributed over the medium; they have a propagation speed of 3000 m/s. Each of the models represents a target that will be normalized and stored in a $256\times256$-array; its respective input consists of the pressure measurements made in 11 fixed positions of the simulated medium, as indicated in \figref{subfig:wave_propagated}. Each measurement starts in the 400th timestep and is stored in a $1400\times11$-array re-scaling them in the range of -50 to 50 units.

\subsection{Convolutional Neural Network Architecture}\label{sec:architecture_CNN}

The proposed CNN has an encoder-decoder architecture, this type of structure consists of two differentiated stages. The first stage is the encoder, it extracts the most relevant features of the input signals and then it encodes them in order to reduce their dimensions. The second stage is called decoder, it interprets the encoded information to produce an output with the desired characteristics. Each of these can be treated as an independent neural network, so in the following sections, we detail each of them.

\begin{figure}[tp]
\subfloat[Encoder]{%
        \centerline{
        \includegraphics[width=0.4\textwidth]{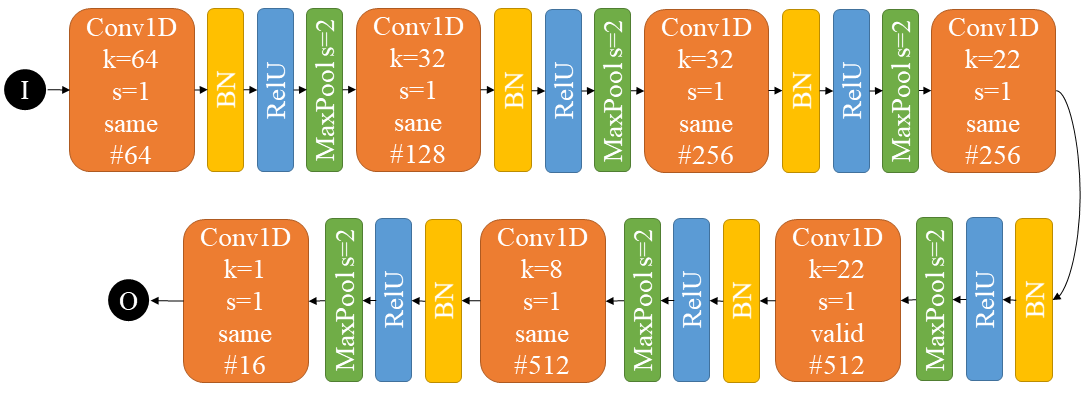}%
        }
        \label{subfig:encoder}
}\\
\subfloat[Decoder]{%
        \centerline{
        \includegraphics[width=0.4\textwidth]{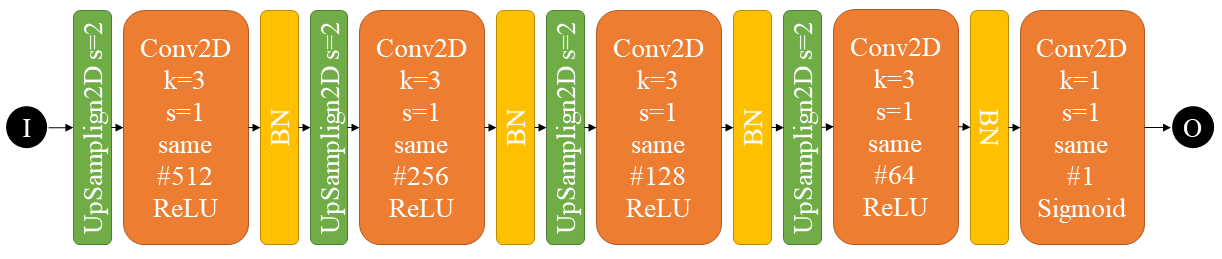}%
        }
        \label{subfig:decoder}
}
\caption{Expanded structure of the encoder and decoder implemented with convolutional layers.}
\label{fig:neural_network}
\end{figure}

\subsubsection{Encoder Structure}\label{sec:encoder_structure}

There is a wide range of encoder structures. Selecting one depends of the shape of the input data and the nature of the phenomenon that produces them \cite{Wu, Sorteberg, Badrinarayanan}. Since our input data are sequences, a traditional approach is using recurrent layers, such as LSTM or GRU, to encode the information; however, the recent use of 1D and 2D convolutional layers have shown great potential to manipulate sequences. In addition to this, they require a lower computational cost, that is why we decided to implement an encoder architecture based on 1D convolutional layers (Conv1D).

The basic structure in the encoder is a sequence of four-layer blocks. Each block is composed by a convolutional layer; followed by a Batch Normalization (BN) layer; an activation layer (ReLU or Sigmoid); and, finally, a max pooling layer. Each layer has its own hyperparameters: kernel size (k) and the padding method (same or valid), for the convolutional layers; window size, for the max-pooling layers; and stride (s), for both. All the hyperparameters of the encoder are shown in \figref{subfig:encoder}.

Regarding the data dimension, the encoder receives an array of $1400\times11$ elements and generates, as an output, an array of $16\times16$ elements. This output contains essential information of the input signals.

\subsubsection{Decoder Structure}\label{sec:decoder_structure}

The decoder structure, in a similar way to the encoder, could be built in many ways, so that its design has to be done considering the nature of both the input and the desired output. Since both input and output of the decoder are two-dimensional arrays, they could be processed as images and, for that reason, 2D convolutional layers (Conv2D) are the main layers in the decoder implementation.

In addition to the Conv2D layers, we also use UpSampling 2D and Batch Normalization (BN) layers. The Conv2D layers have almost the same hyperparameters as the Conv1D layers with the only difference that their kernels are two-dimensional. Moreover, the Up Sampling layers have two hyperparameters: window size and stride, which have values equal to 2 in the decoder implementation. The hyperparameters values for each layer are shown in \figref{subfig:decoder}.

\section{Results}\label{sec:results}

In this section, we will show the training results of the CNN models, as well as some metrics used in the process to evaluate the their performance.

\subsection{CNN Training}\label{sec:training_stage}

\begin{figure}[!t]
    \centerline{
    \includegraphics[width=0.35\textwidth]{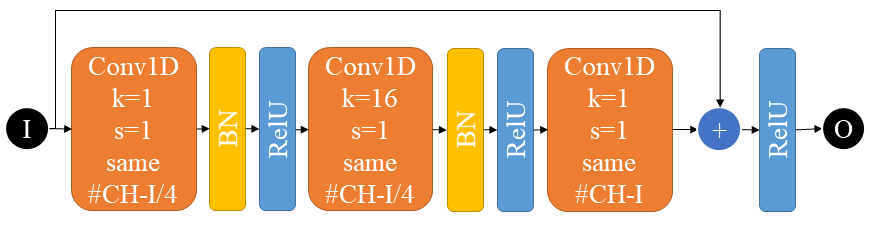}}
    \caption{Residual Layer}
    \label{fig:residual_layer}
\end{figure}

In this stage, we propose two additional models in order to compare their performance with that of the model proposed in \secref{sec:architecture_CNN}. These models have residual layers, similar to that shown in \figref{fig:residual_layer}. From now on, the CNN described in \secref{sec:architecture_CNN} will be called InvNet, while the additional models will be called InvNet+1Res and InvNet+2Res; where, notation +1Res and +2Res are references to the number of residual layers added after each max pooling layer in the InvNet's encoder.

\begin{figure}[!b]
\centering
\subfloat[Accuracy]{%
        \includegraphics[clip,width=0.4\columnwidth]{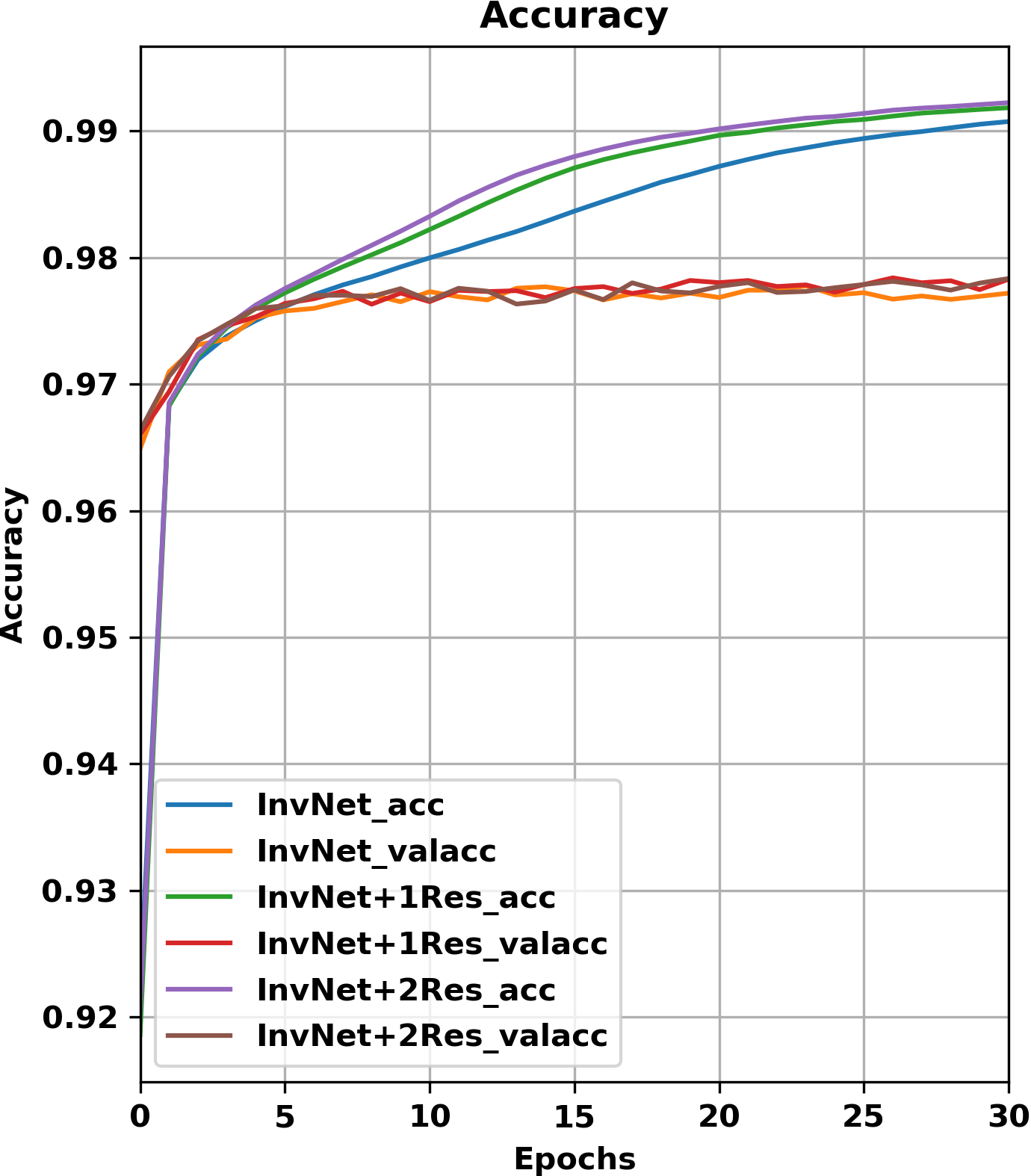}%
        \label{subfig:accuracy}
}
\subfloat[Loss]{%
        \includegraphics[clip,width=0.4\columnwidth]{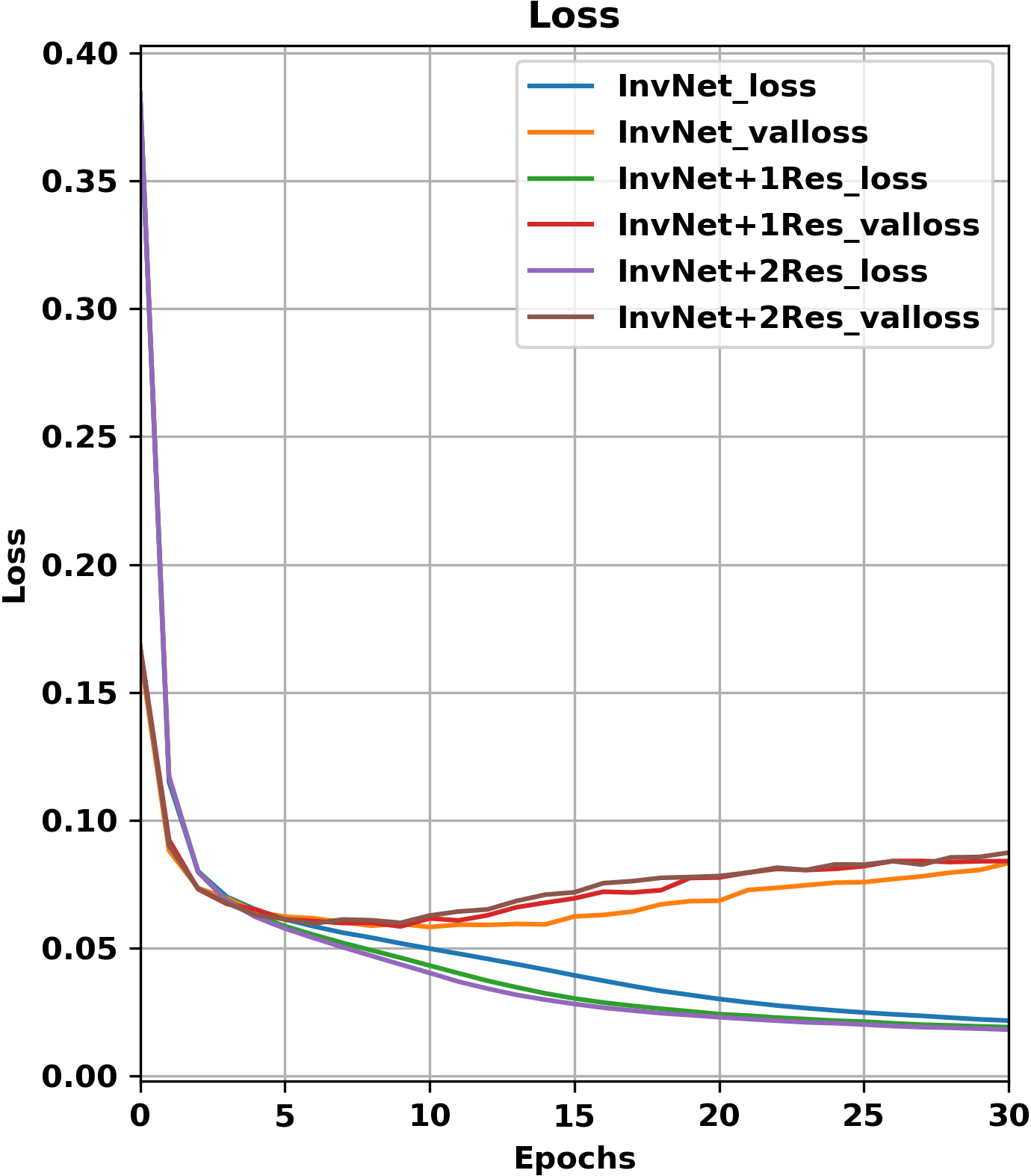}%
        \label{subfig:loss}
}
\caption{Curves obtained during the training stage of the three CNNs.}
\label{fig:curves}
\end{figure}

The proposed CNNs were implemented with Python 3.6 in a server with an Intel Xeon E5-2620 CPU at 2.1 GHz, 128GB RAM and two Nvidia Tesla K40 GPU. The dataset described in \secref{sec:dataset_generation} was divided in training, validation and test sets with a ratio of 70~\%, 15~\% and 15~\%, respectively. Then, the selected cost function was binary cross-entropy, and the optimizer was Adam with a learning rate equal to 0.0002, a first moment ($\beta_1$) equal to 0.5 and a second moment ($\beta_2$) equal to 0.99. Finally, all CNNs were trained with a batch size of 20 during 30 epochs; accuracy and loss curves are shown in \figref{fig:curves}, where it could be seen that they tend to over-fit around the 10th epoch. In the validation set, the CNNs get accuracy and loss values around 97.6~\% and 0.064, respectively.

\subsection{Evaluation of the CNN}\label{sec:evaluation_stage}

After training, we proceed to evaluate and compare the performance achieved by each model. Since the outputs of the CNNs are binary masks, the following metrics will be used in the analysis: accuracy, precision, sensitivity, specificity and intersection over union (IoU).

The accuracy, precision, sensitivity and specificity are calculated with \eqref{eq:accuracy}, \eqref{eq:precision}, \eqref{eq:recall} and \eqref{eq:specificity}; where $tp$ is the number of true positive pixels; $tn$, true negatives; $fp$, false positives; and $fn$, false negatives. Each one of this are measured in relation with pixels of the output image. While accuracy is a global metric that indicates the percentage of pixels correctly classified as solid objects or water, the precision indicates the percentage of pixels properly detected over objects. In a similar way, the sensitivity points out the percentage of pixels of the objects that have been omitted and the specificity is related with the percentage of pixels which are correctly classified as water. 

Additionally, IoU measures the percentage of overlap between the ground truth and the estimated velocity model, calculated according to \eqref{eq:IoU}. It gives us an intuition of how well located and sized the objects are.
\begin{figure}[tp]
\centering
\subfloat[][]{%
        %\rule{4cm}{3cm}
        \includegraphics[clip,width=0.27\columnwidth]{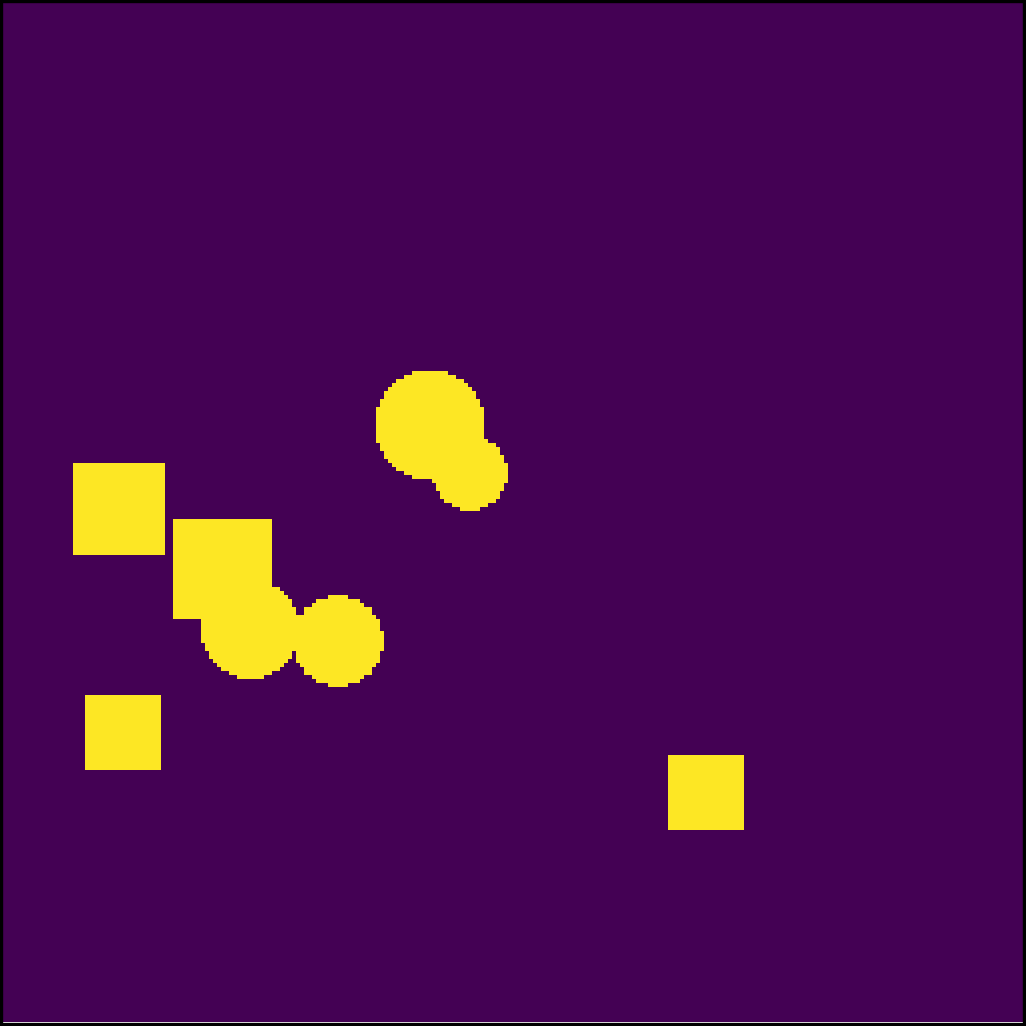}%
        \label{subfig:gt1}
}
\subfloat[][]{%
        %\rule{4cm}{3cm}
        \includegraphics[clip,width=0.27\columnwidth]{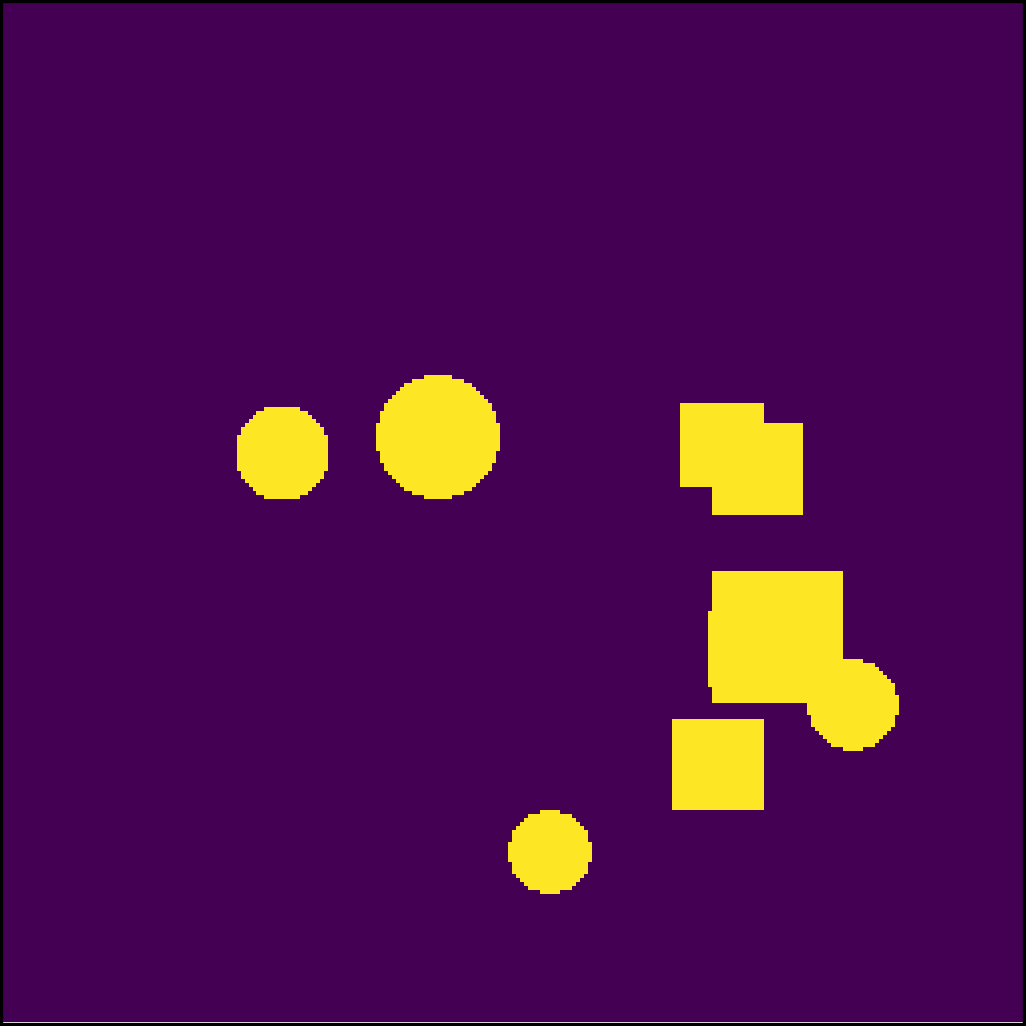}%
        \label{subfig:gt2}
}
\subfloat[][]{%
        %\rule{4cm}{3cm}
        \includegraphics[clip,width=0.27\columnwidth]{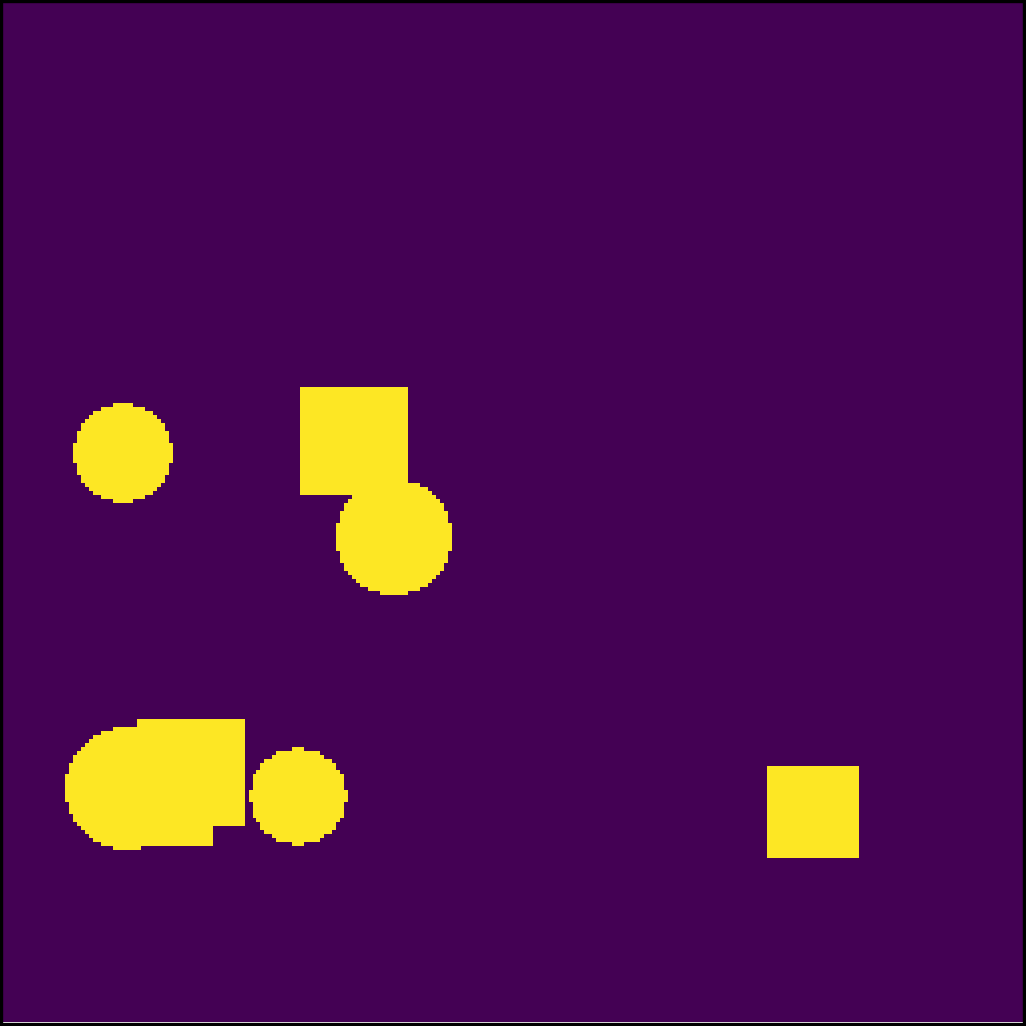}%
        \label{subfig:gt3}
}\\
\subfloat[][]{%
        %\rule{4cm}{3cm}
        \includegraphics[clip,width=0.27\columnwidth]{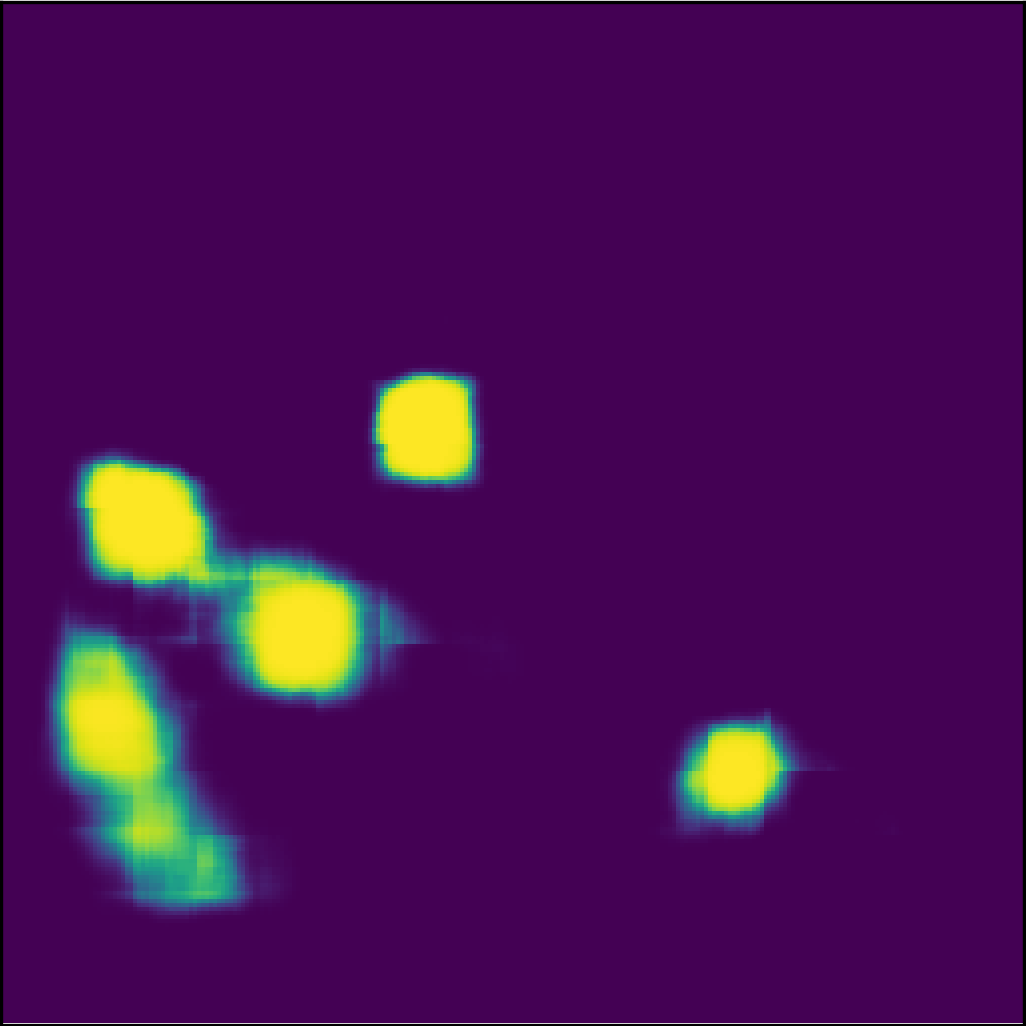}%
        \label{subfig:es1}
}
\subfloat[][]{%
        %\rule{4cm}{3cm}
        \includegraphics[clip,width=0.27\columnwidth]{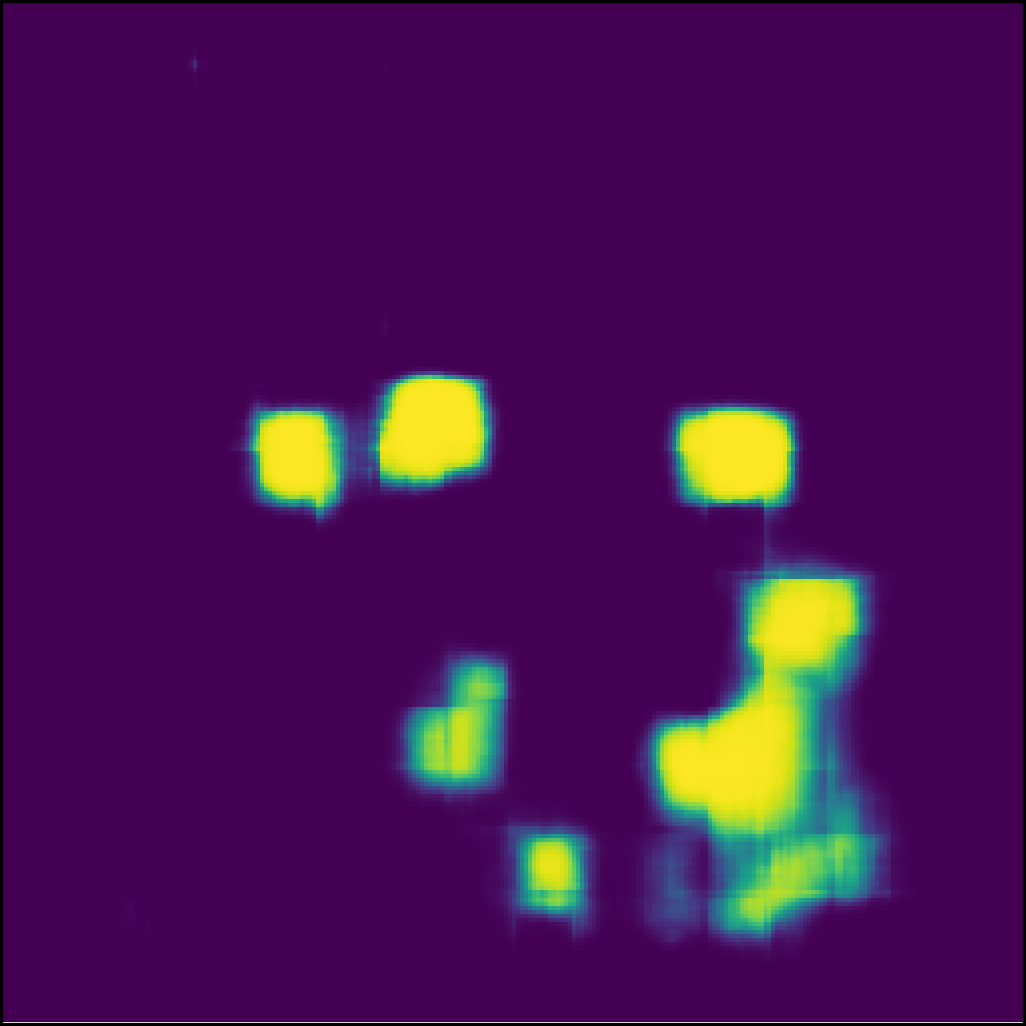}%
        \label{subfig:es2}
}
\subfloat[][]{%
        %\rule{4cm}{3cm}
        \includegraphics[clip,width=0.27\columnwidth]{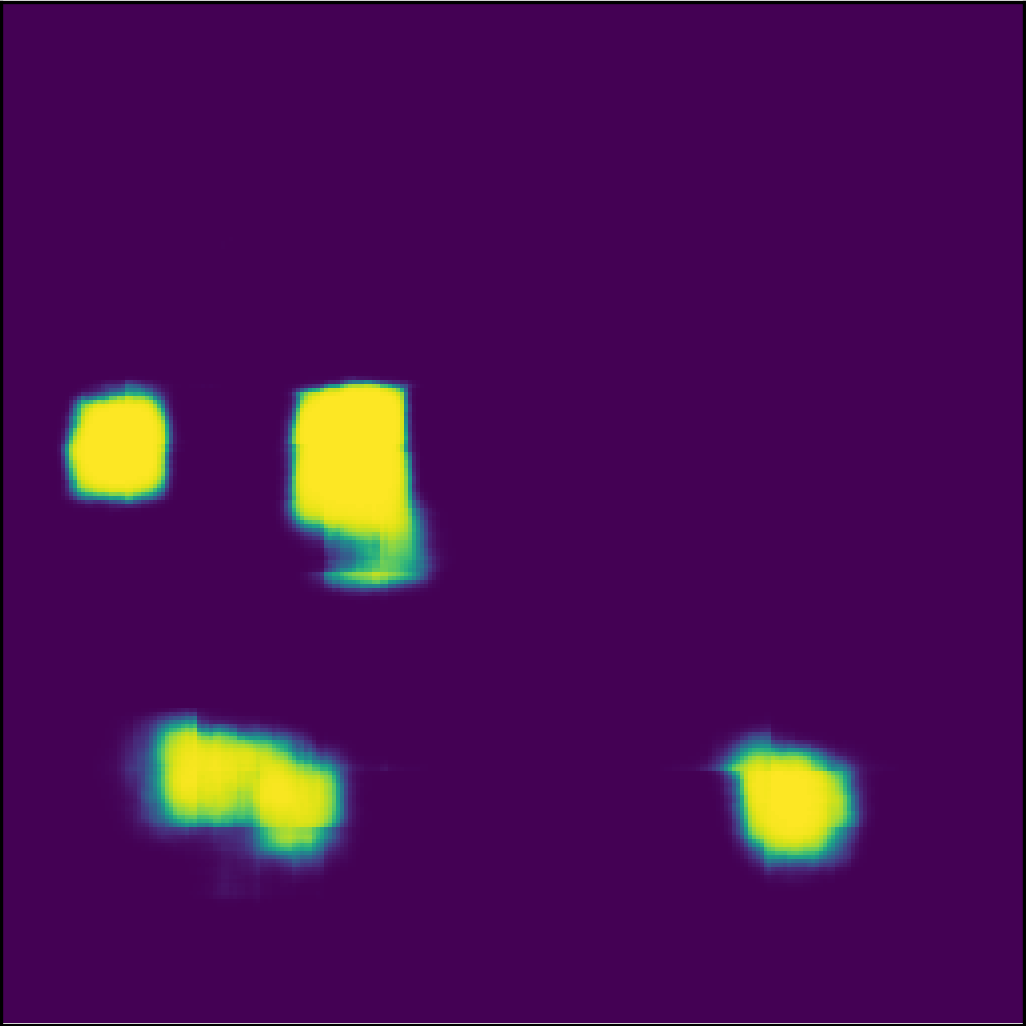}%
        \label{subfig:es3}
}
\caption{Images (a), (b) y (c) show the ground truth velocity model, while images in right side (d), (e) y (f) show the model estimated by InvNet.}
\label{fig:evaluation}
\end{figure}

\begin{equation}
\label{eq:accuracy}
accuracy = (tp + tn) / (tp + fp + tn + fn)
\end{equation}
\begin{equation}
\label{eq:precision}
precision = tp / (tp + fp)
\end{equation}
\begin{equation}
\label{eq:recall}
sensitivity = tp / (tp + fn)
\end{equation}
\begin{equation}
\label{eq:specificity}
specificity = tn / (tn + fp)
\end{equation}
\begin{equation}
\label{eq:IoU}
IoU = (target \cap prediction) / (target \cup prediction)
\end{equation}

These metrics were used over the test set, which has 3000 samples, for each CNN and the results are shown in \tabref{t:metrics}. There, it could be seen that all the CNNs have high values in the accuracy and specificity metrics, but lower values in the precision and sensitivity metrics. These results point out that they can detect the presence of objects but they still have difficulty interpreting interference and echoes. In the case of the IoU metric, all the CNNs obtain high values which indicates that all of them can properly estimate the localization and size of the objects. Despite the InvNet+1Res have a slightly higher performance than the others, it has a higher computational cost than that of InvNet, based on the required number of parameters shown in \tabref{t:metrics}. Since, InvNet has a good performance and low computational cost, it is the best of the three models.

\begin{table}[!b]
\caption{Performance obtained from the CNNs trained}
\begin{center}
\begin{tabular}{|l | l| l| l|}
    \hline
    \textbf{Models}&\textbf{InvNet}&\textbf{InvNet+1Res}&\textbf{InvNet+2Res} \\
    \hline
    \textbf{Accuracy} & 97.730\% & 97.853\% & 97.848\% \\
    \hline
    \textbf{Precision} & 75.880\% & 76.444\% & 76.359\% \\
    \hline
    \textbf{Sensitivity} & 64.692\% & 67.323\% & 67.285\% \\
    \hline
    \textbf{Specificity} & 99.131\% & 99.134\% & 99.130\% \\
    \hline
    \textbf{IoU} & 98.589\% & 98.718\% & 98.708\%\\
    \hline
    \textbf{Parameters} & 9M & 10M & 11M\\
    \hline
\end{tabular}
\label{t:metrics}
\end{center}
\end{table}

Finally, we test InvNet under different situations and the results are shown in \figref{fig:evaluation}. There, it could be seen what was mentioned above; that is, the proposed CNN correctly locates most of the objects, it also estimates their shapes and sizes, but presents some false positives as observed in \figref{subfig:gt2} and \figref{subfig:es2}, as well as some omissions of objects as seen in \figref{subfig:gt3} and \figref{subfig:es3}.

\section{Conclusions}\label{sec:conclusion}

In this work, a convolutional encoder-decoder architecture was proposed to estimate the velocity model of an underwater environment, managing to locate objects and approximate their shapes with a high value in the  IoU metric equals to 98.58\%. Despite the fact that the CNN presents a good performance, it can still be improved for in some cases it makes mistakes in detecting objects due to multiple echoes and shadows. These behaviors reflect their effects in the precision and sensitivity metrics where the CNN obtains values of 75.88\% and 64.69\%, respectively. 

Additionally, it is important to point out that the proposed CNN shows characteristics such as quick calculation of a velocity model based on acoustic signals, and precision in objects localization and size estimation. Since the proposed model was trained in a great variety of synthetic scenarios, we can infer that it may achieve similar results with signals from real scenarios. This is part of a future work.

%Although, the model proposed here was trained using only synthetic data, it yet have showed some promising features to highlight   

\section*{Acknowledgment}

The authors would like to thank the National Institute for Research and Training in Telecommunications (INICTEL-UNI) for the technical and financial support to carry out this work.

\end{document}